\documentclass[%
 reprint,
 amsmath,amssymb,
 aps,
]{revtex4-1}

\usepackage{amsmath}
\usepackage{amssymb}
\usepackage{epsfig}
\usepackage{graphicx}
\usepackage{dcolumn}
\usepackage{bm}

\begin{document}

\preprint{APS/123-QED}

\title{A sufficient condition of violating the SPA conjecture\\}

\author{Bang-Hai Wang$^{1,2}$ and Dong-Yang Long$^2$}%

\affiliation{%
$^1$School of Computers, Guangdong University of Technology, Guangzhou 510006, People's Republic of China
}%
\affiliation{%
$^2$Department of Computer Science, Sun Yat-sen University, Guangzhou 510006, People's Republic of China
}%

\date{ \today}

\begin{abstract}
Based on the general form of entanglement witnesses constructed from separable states, we first show a sufficient condition of violating the structural physical approximation (SPA) conjecture [\emph{Phys. Rev. A }{\bf 78,} 062105 (2008)]. Then we discuss the SPA conjecture for decomposable entanglement witnesses. Moreover, we make geometric illustrations of the connection between entanglement witnesses and the sets of quantum states, separable states, and entangled states comparing with planes and vectors in Euclidean space.

\end{abstract}
\maketitle

\section{Introduction}

Quantum entanglement is considered as the central resource for quantum information processing \cite{Horodecki09,Guhne09}, such as quantum computation, quantum dense coding, quantum teleportation, quantum cryptography, etc. However, quantum entanglement is still not full known by researchers. It is one of the main research topics in the theory of entanglement that how to detect a given state entangled or separable.

To the best of our knowledge, positive (linear) maps \cite{M.Horodecki96} up to date may be the most powerful method among various separability criteria. For any entangled state, there exists at least a positive but not completely positive map to detect it. As a consequence of the Jamio{\l}kowski-Choi isomorphism \cite{Jam}, entanglement witnesses (EWs) \cite{Terhal00} play the role to detect entanglement equivalently. An observable $W=W^\dagger$ is said an EW if (i) the expectation value of $W$ is non-negative for any separable state; and (ii) it is negative for at least an entangled state. Naturally, (iii) $\gamma W$ keeps all properties of $W$ as an EW for a non-negative number $\gamma$. In this case, we say that $\gamma W$ is the same EW as $W$. An EW which detects a maximal set of entanglement is defined to be optimal in Ref. \cite{Lewenstein00}. An EW whose expectation value vanishes on at least one product vector is said to be weakly optimal \cite{Badziag07}. A necessary condition for an EW $W$ to be optimal is that there must exist a separable $\sigma$ with $\text{tr}(W\sigma)=0$ \cite{Guhne09}. It is, however, a sufficient condition for the weakly optimal EW (WOEW) but not for the optimal EW (OEW). We say $W_2$ is finer than $W_1$ if the entangled state detected by $W_2$ is more than the one by $W_1$ \cite{Lewenstein00}.

For our purpose, we can only consider the quantum states on the finite
dimensional Hilbert space
$\mathcal{H}_{AB}=\mathcal{H}_A\otimes\mathcal{H}_B$. We let
dim$(\mathcal{H}_{A})={d}_{A}$, dim$(\mathcal{H}_{B})={d}_{B}$ and
dim$(\mathcal{H}_{AB})={d}_{AB}$.

\section{The entanglement witnesses and structural physical approximations}

Our recent work \cite{Wang11} showed that any EW $W$ can be written as
\begin{equation}
W=\sigma-c_\sigma I\,\label{ew-form0}
\end{equation} where $\sigma$ is a
(normalized) separable density matrix and $\lambda_{0\sigma}<c_\sigma\leq c_\sigma^{max}$ is a real number related to $\sigma$, $\lambda_{0\sigma}$ is the minimum
eigenvalue of $\sigma$ and
\begin{equation}
c_\sigma^{max}=\inf_{\parallel|\mu_A\rangle\parallel=1,
\parallel|\nu_B\rangle\parallel=1}\langle\mu_A\nu_B|\sigma|\mu_A\nu_B\rangle
\end{equation}
is the maximum number in
$c_\sigma$ which makes $W=\sigma-c_\sigma I$ an EW and $|\mu_A\nu_B\rangle$ is any unit product vector. Clearly, an EW $W=\sigma-c_{\sigma} I$ is weakly optimal if and only if $c_{\sigma}=c_\sigma^{max}$, and there exists a separable density matrix $\sigma$ with $W^{opt}=\sigma-c_\sigma^{max}I$ for any OEW $W^{opt}$.

\subsection{A sufficient condition of violating the SPA conjecture}

Recently, a conjecture that SPA to optimal positive maps correspond to entanglement-breaking maps (channels) has been posed in \cite{Korbicz08}. An equivalent
presentation of the conjecture (called SPA conjeture) is that SPA to optimal entanglement witnesses correspond to separable (unnormalized) states, i.e., the SPA
to an optimal EW $W^{opt}$,
\begin{eqnarray}
 \tilde{W}^{opt}=W^{opt}+sI\label{SPAstatetogeneral},
\end{eqnarray} where $s>0$ is the smallest parameter for
which $\tilde{W}$ is a positive operator (possibly unnormalized state) \cite{Wang11}.

Based on this result, we found that the SPA conjecture does not depend on the optimality of EWs. We have found that the separability problems of SPA to both optimal EWs and non-optimal EWs become
 the same problem, that is, whether any
 \begin{equation}
 \tilde{W}=\sigma-\lambda_{0\sigma}
 \end{equation}
  is separable with $W=\sigma-c_\sigma I$ being an EW ($W=\sigma-c_\sigma^{max} I$
 being an OEW) \cite{Wang11}.

\textbf{Theorem 1.} If $W=\sigma-c_\sigma I$ is an EW with $\sigma$ being not full rank, its SPA defines an entanglement-breaking channel (EBC) (the output is just $\sigma$).

Unless otherwise specified, EWs with $W=\sigma-c_\sigma I$ discussed below refer to EWs with $\sigma$ being full rank.

Following the definition in Ref. \cite{Lewenstein00}, if an EW can be written in the form $W=P+Q^\Gamma$ with $P, Q\geq0$, we say it decomposable, otherwise we say it indecomposable. It is well known that the division of EWs to
decomposable and indecomposable is translated from positive maps via the Jamio{\l}kowski-Choi isomorphism \cite{Jam}.

For simplicity, we consider optimal nondecomposable EWs (ONEWs).

\textbf{Lemma 1 \cite{Lewenstein00,Korbicz08}.} $W$ is an ONEW if and only if $W^\Gamma$ is an optimal nondecomposable EW, where $\Gamma$ denotes the partial transposition.

\textbf{Theorem 2.} If there exist an ONEW $W=\sigma-c_\sigma^{max}I$ with $\lambda_{0\sigma^\Gamma}<\lambda_{0\sigma}$, $W$ violates the SPA conjecture.

\textbf{Proof:} By the SPA of $W=\sigma-c_\sigma^{max}I$, $\tilde{W}=\sigma-\lambda_{0\sigma}I$, $\tilde{W}^\Gamma=\sigma^\Gamma-\lambda_{0\sigma}I$. By $\lambda_{0\sigma^\Gamma}<\lambda_{0\sigma}$, $\tilde{W}^\Gamma=\sigma^\Gamma-\lambda_{0\sigma}I<0$, and $\tilde{W}=\sigma-\lambda_{0\sigma}I$ is not separable by positive partial transposition (PPT) criterion \cite{Peres96}. \hfill$\blacksquare$

\textbf{Corollary 1.} If $W=\sigma-c_\sigma^{max}I$ is an ONEW with $\lambda_{0W^\Gamma}\neq \lambda_{0W}$, the SPA of $W$ or the SPA of $W^\Gamma$ violates the SPA conjecture.

Following Corollary 1, the SPA conjecture is not true if there exists an EW with its minimum
eigenvalue being not equal to the minimum eigenvalue of its partial transposition. Since the result of Theorem 2 is followed with the PPT criterion and the PPT criterion is a necessary but not a sufficient
separable condition for separability, it is not easy to find the necessary condition of violating the SPA conjecture by our result.

\subsection{An example and discussion}

Since all structural approximations to positive maps of low dimensions define entanglement-breaking channels \cite{Wang11}, it is not easy to construct the counterexample of the SPA conjecture. Very recently, Ha and Kye \cite{Ha12} exhibited counterexamples of indecomposable EWs violating the SPA conjecture. St{\o}rmer \cite{Stormer12} gave a sufficient condition of violating the SPA conjecture by their theory.

Consider the ONEW in \cite{Ha12}
\begin{equation}
W[a,b,c;\theta]= \left( \begin{array}{ccccccccccc} a     &\cdot   &\cdot  &\cdot  &-e^{i\theta}     &\cdot
&\cdot   &\cdot  &-e^{-i\theta}     \\ \cdot   &c &\cdot    &\cdot    &\cdot   &\cdot &\cdot &\cdot     &\cdot   \\ \cdot  &\cdot    &b &\cdot &\cdot  &\cdot
&\cdot    &\cdot &\cdot  \\ \cdot  &\cdot    &\cdot &b &\cdot  &\cdot    &\cdot    &\cdot &\cdot  \\ -e^{-i\theta}     &\cdot   &\cdot  &\cdot  &a     &\cdot
&\cdot   &\cdot  &-e^{i\theta}     \\ \cdot   &\cdot &\cdot    &\cdot    &\cdot   &c &\cdot &\cdot    &\cdot   \\ \cdot   &\cdot &\cdot    &\cdot    &\cdot
&\cdot &c &\cdot    &\cdot   \\ \cdot  &\cdot    &\cdot &\cdot &\cdot  &\cdot    &\cdot    &b &\cdot  \\ -e^{i\theta}     &\cdot   &\cdot  &\cdot  &-e^{-i\theta}
&\cdot   &\cdot &\cdot  &a \end{array} \right).
\end{equation}

Let $\theta=\pi/12$, $a=\frac{4}{3}cos\frac{\pi}{12}$, $b=\frac{2}{3}cos\frac{\pi}{12}$, and $c=0$. We compute
$\lambda_{0W^\Gamma}\approx-0.6440, \lambda_{0W}\approx-0.7286$, and $\lambda_{0W^\Gamma}\neq\lambda_{0W}$. By Corollary 1, $W[\frac{4}{3}cos\frac{\pi}{12},\frac{2}{3}cos\frac{\pi}{12},0;\pi/12]$ violates
the SPA conjecture.

Let $W=\sigma-c_\sigma^{max}I$ be an ONEW. Suppose the spectral decomposition
of $\sigma^\Gamma$, $\sigma^\Gamma=\sum_{i=0}^{d_{AB}-1}\lambda_{i\sigma^\Gamma}|e_i\rangle\langle e_i|$, the spectral decomposition
of $\sigma$, $\sigma=\sum_{i=0}^{d_{AB}-1}\lambda_{i\sigma}|f_i\rangle\langle f_i|$. $W=\sigma-c_\sigma^{max}I$ is an EW of $|f_0\rangle\langle f_0|$ and $|e_0\rangle\langle e_0|^\Gamma$; $W^\Gamma=\sigma^\Gamma-c_\sigma^{max}I$ is an EW of $|f_0\rangle\langle f_0|^\Gamma$ and $|e_0\rangle\langle e_0|$.

It is well known that decomposable EWs cannot detect PPT entangled states.  If $W=\sigma-c_\sigma^{max}I$ is a decomposable OEW, $W^\Gamma=Q\geq0$ with $Q$ supporting by an entangled subspace and $W^\Gamma$ is entangled. Moreover, we have $\lambda_{0\sigma}<\lambda_{0\sigma^\Gamma}$ and $\tilde{W}^\Gamma=\sigma^\Gamma-\lambda_{0\sigma}I>0$ for the decomposable EW $W=\sigma-c_\sigma I$.

Clearly, if $\tilde{W}=\sigma-\lambda_{0\sigma}I$ is entangled with $W=\sigma-c_\sigma I$ being a decomposable EW, it is a PPT entangled state and there exists an indecomposable EW detect it. If $W=\sigma-c_\sigma I$ is a decomposable EW with the spectral decomposition $\sigma=\sum_i\lambda_{i\sigma}|f_i\rangle\langle f_i|$, $\pi=|f_0\rangle\langle f_0|>0$ is entangled and is not a PPT entangled state, and one of its EW is $W=\sigma-c_\sigma I$, $\pi^\Gamma=|f_0\rangle\langle f_0|^\Gamma$ is an EW \cite{Wang11}. By $\pi^\Gamma=|f_0\rangle\langle f_0|^\Gamma$ is optimal \cite{Augusiak11} and $tr(\tilde{W}^\Gamma\pi^\Gamma)=tr(\tilde{W}\pi)=0$. Although we cannot obtain that $\tilde{W}$ is separable by $tr(\tilde{W}^\Gamma\pi^\Gamma)=0$ for the optimal EW $\pi^\Gamma$, it seems that the SPA conjecture is true for decomposable EWs as stated in \cite{Ha12}.


\section{Geometric illustrations of entanglement witnesses and quantum states}

Considering the finite-dimensional Hilbert space $\mathcal{H}_{AB}=\mathcal{C}^{d^2_{AB}}$ as in \cite{Bertlmann02}, where observables $W$ are represented by all Hermitian matrices and states $\rho$ by density matrices, we can regard these quantities as elements of a real Hilbert space $\mathcal{H}_r=\mathcal{R}^{d^2_{AB}}$ with scalar product

\begin{equation}
\langle\rho|W\rangle=tr\rho W
\end{equation}
and corresponding norm
\begin{equation}
\|W\|_2=(trW^2)^{\frac{1}{2}}.
\end{equation}
Both density matrices with trace unity and observables are represented by vectors in $\mathcal{H}_r$.

By Hahn-Banach theorem, we can view an EW $W$ as a hyperplane, which has dimension $d_{AB}-1$ \cite{Note1}. We can view $W$ as a separating hyperplane with entangled state $\pi$ on one side and all separable states on the other\cite{Pittenger02}. We can rewrite Eq. (\ref{ew-form0}) as
\begin{equation}
W=\sigma-c'_\sigma\tau_0,\label{ew-density0-sub}
\end{equation}
where $\tau_0=\frac{1}{d_{AB}}I$ denotes the maximally mixed state (separable). Therefore, any EW is equal to the difference between a separable state $\sigma$ and the product of a non-negative real number and the maximally mixed state. These can be compared with the case of Euclidean space. A plane is defined by its orthogonal vector in Euclidean space. The plane separates vectors having a negative scalar product with the orthogonal vector from vectors having a positive one. Vectors in the plane have a vanishing scalar product with the orthogonal vector. A vector can add or subtract another vector in Euclidean space.

\begin{figure}
\epsfig{file=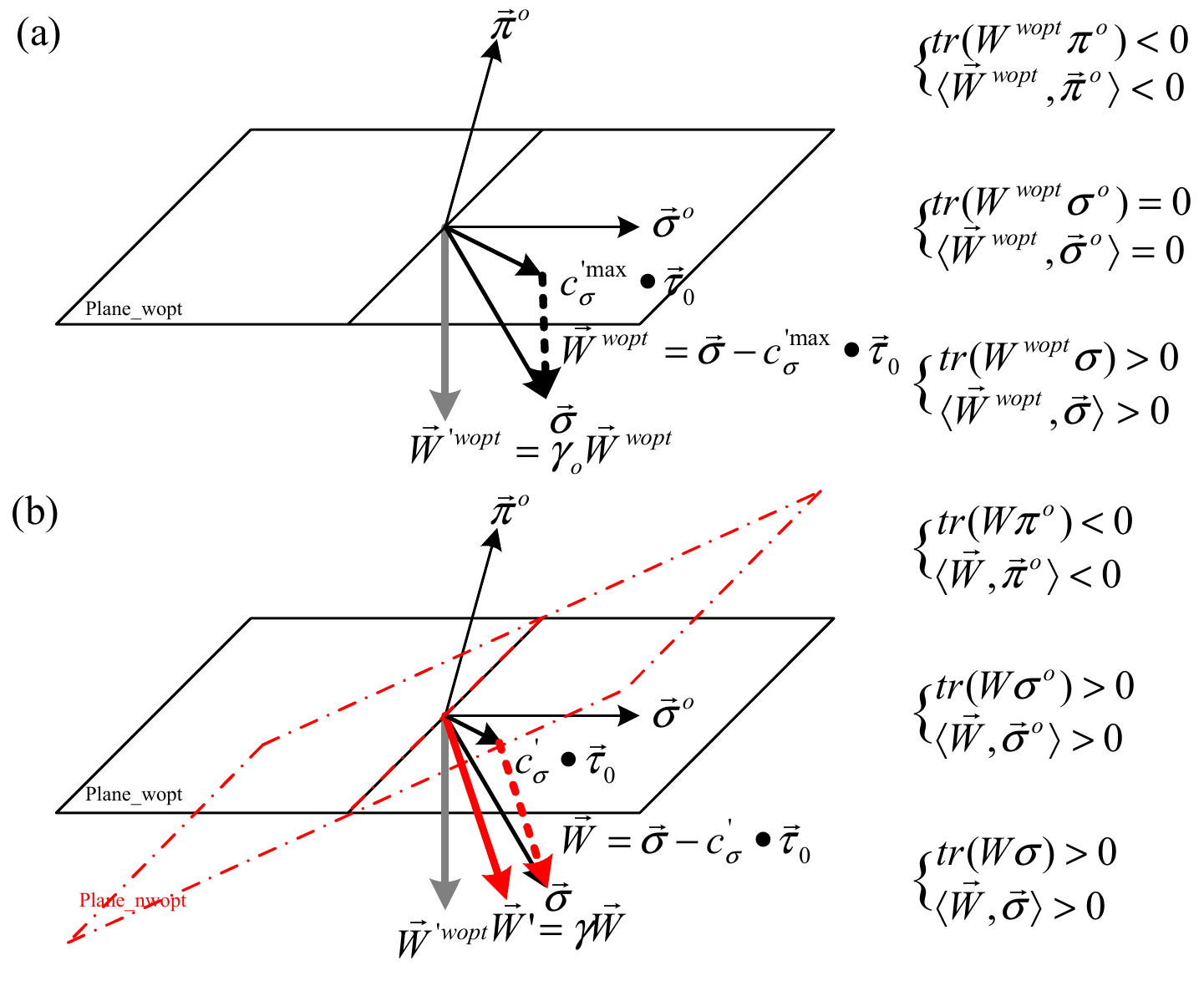,width=.95\columnwidth}
  \caption{Geometric illustration of (a) The weakly optimal entanglement witness $W^{wopt}=\sigma-c'^{max}_\sigma\tau_0$ and the ``weakly optimal'' plane $\overset{\lower0.5em\hbox{$\smash{\scriptscriptstyle\kern-16pt\rightarrow}$}}{W^{wopt}}$; (b) The non-weakly-optimal and weakly optimal entanglement witnesses $W^{wopt}$, $W^{nwopt}=\sigma-c'_\sigma\tau_0$ and the ``non-weakly-optimal'' and ``weakly-optimal'' planes $\overset{\lower0.5em\hbox{$\smash{\scriptscriptstyle\kern-16pt\rightarrow}$}}{W^{nwopt}}$, $\overset{\lower0.5em\hbox{$\smash{\scriptscriptstyle\kern-16pt\rightarrow}$}}{W^{wopt}}$.
  }\label{fig1}
\end{figure}

We have a geometrical illustration of the weakly optimal EW $W^{wopt}$, a non-weakly-optimal EW $W$, and quantum states comparing planes and vectors in Euclidean space as in \cite{Bertlmann05}, as shown in Fig. 1. Hyperlanes separate ``left-hand" entangled states from ``right-hand" separable states. $W^{wopt}=\sigma-c'^{max}_\sigma\tau_0$ is the same weakly optimal EW as $W'^{wopt}=\gamma_o W^{wopt}$, where $\sigma$ is a separable state, $\tau_0$ is the maximally mixed state, and $c'^{max}_\sigma$ and $\gamma_o$ are non-negative numbers. $\vec{W}^{wopt}=\vec{\sigma}-c'^{max}_\sigma\vec{\tau_0}$ denotes the same plane as the parallel $\vec{W}^{'wopt}$. $\pi^o$ is entangled such that tr$(W^{wopt}\pi^o)<0$. $\sigma^{o}$ inside the weakly optimal hyperlane is separable such that tr$(W^{wopt}\sigma^{o})=0$.

The relation between separable states, entangled states and different EWs in the form $W=\sigma-c_\sigma I$ is shown in Fig. 2. EWs are denoted as planes in Hermitian operator space. Sets $Q$, $S$, and $E$ of quantum states, separable states, and entangled states are such that $Q=S\cup E$. $W^{wopt}=\sigma-c_\sigma^{max} I$ and $W^{'wopt}=\sigma'-c_\sigma'^{max} I$ are weakly optimal entanglement witnesses. $W_2=\sigma-c_{2\sigma} I$ is finer than $W_1=\sigma-c_{1\sigma} I$. The ``boundary" of EWs, $\bar{W}=\sigma-\lambda_{0\sigma} I$ is not an entanglement witness, but a quantum state, entangled or separable.

\begin{figure}
\epsfig{file=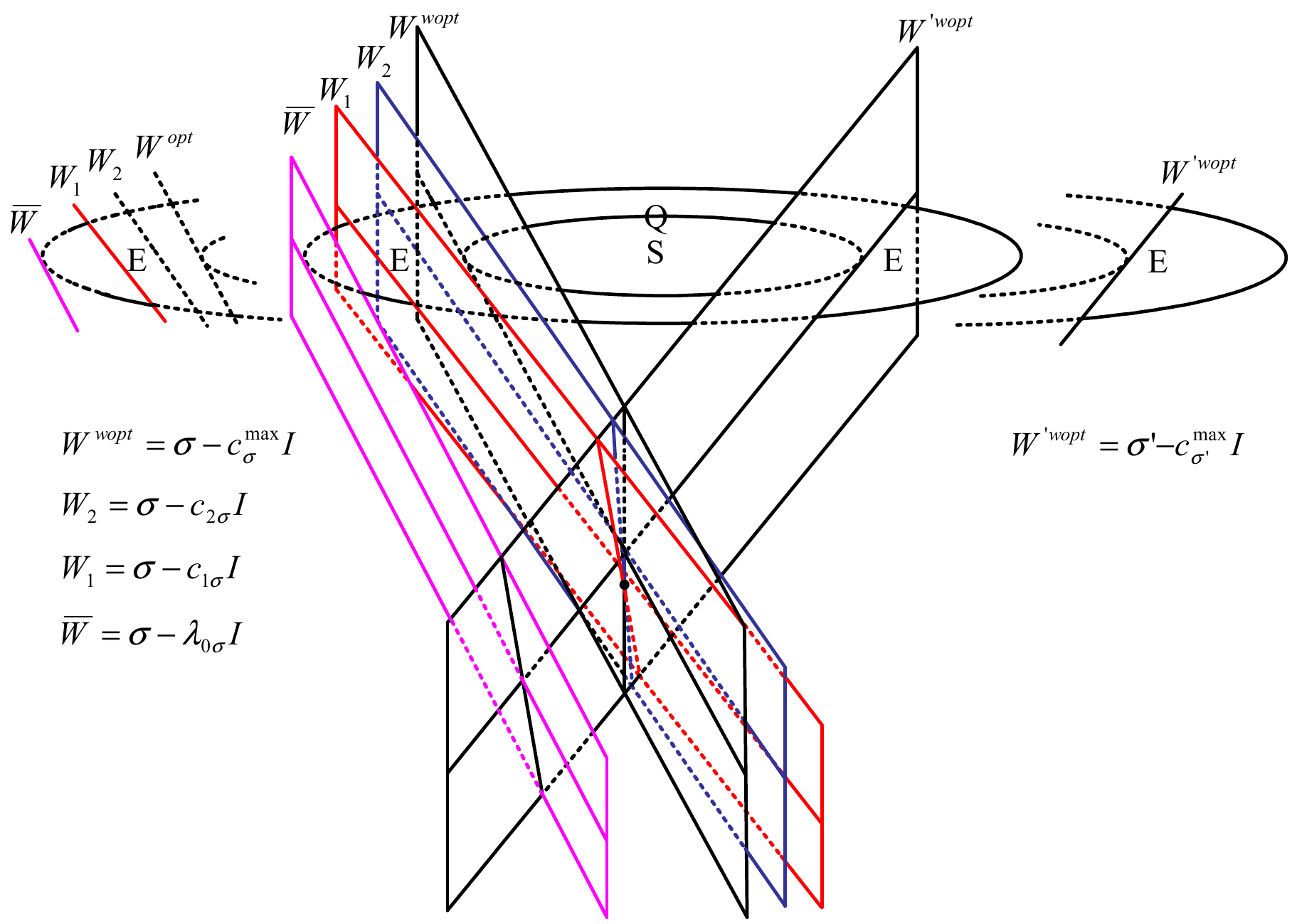,width=.95\columnwidth}
  \caption{By the general form of EWs constructed from separable states, schematic representation of entanglement witnesses and sets of quantum states, separable states, and entangled states. }\label{fig2}
\end{figure}

\section{Summary}

In summary, we give a sufficient condition of
violating the SPA conjecture [\emph{Phys. Rev. A }{\bf 78,} 062105 (2008)] following the general form of EWs constructed from sparable states. Comparing with planes and vectors in Euclidean space, we make geometric illustrations of the connection between entanglement witnesses and the sets of quantum states, separable states, and entangled states.

\begin{acknowledgments} We would like to thank Professor Guang Ping He, Rui-Gang Du, Ning-Yuan Yao, Dan Wu and A. Winter for helpful discussions and suggestions. This work was supported by the National Natural Science Foundation of China under Grants No.61272013 and the National Natural Science Foundation of Guangdong province of China under Grants No.s2012040007302. \end{acknowledgments}

\end{document}